Switching magnetoresistance in vertically interfaced $Pr_{0.5}Ca_{0.5}MnO_3$ grown on ZnO nanowires

R. V. K. Mangalam,[1] Z. Zhang,[2] T.Wu,[2] and W. Prellier[1*]

[1]Laboratoire CRISMAT, ENSICAEN, CNRS UMR 6508,
6 Boulevard Maréchal Juin, F-14050 Caen Cedex, France
[2]Division of Physics and Applied Physics, School of Physical
and Mathematical Sciences, Nanyang Technological
University, Singapore 637371

The synthesis, morphology and magneto-transport properties of nanostructure-engineered charge-ordered $Pr_{0.5}Ca_{0.5}MnO_3$ grown on ZnO nanowires are reported. The stability of the charge-ordering can be tuned, but more interestingly the sign of the magnetoresistance is inverted at low temperatures. Coexistence of ferromagnetic clusters on the surface and antiferromagnetic phase in the core of the grains were considered in order to understand these features. This work suggests that such a process of growing on nanowires network can be readily extended to other transition metal oxides and open doors towards tailoring their functionalities.

PACS numbers: 81.15.Fg, 75.47.Lx, 73.63.-b, 75.25.Dk

Perovskite manganites exhibits a wide variety of physical phenomena such as colossal magnetoresistance, charge-ordering, and phase transition, etc.[1] These physical phenomena can be tuned internally by ion replacement and externally by magnetic field, temperature and pressure. In thin films, it can be manipulated further by strain engineering.[2] $Pr_{0.5}Ca_{0.5}MnO_3$ (PCMO) is a well known antiferromagnetic (AFM) charge-ordered (CO) system.[3] The electrons localization due to the ordering of $Mn^{3+}$ and $Mn^{4+}$ ions on specific lattice sites in PCMO leads to the CO state. Below CO transition temperature ($T_{CO}$), PCMO becomes insulating, and this insulating state can be suppressed under application of high magnetic field,[4] though this field is reduced when the materials is deposited as thin films.[5] Recently, it was shown that the charge-ordered state can also be suppressed in nanocrystalline PCMO by modification of the grain size.[6] Additionally, the surface effects due to lower grain size leads, in such cases, to a ferromagnetic (FM) state at the surface and AFM/CO phase in the core. The origin of this FM behavior in nanomaterials is still a subject of debate since it can also be a result of electron transfer to the cations at the surface by the oxygen deficiency.[7] Also, it would become even more complex in the case of materials such as phase-separated manganites whose ground state itself can be FM or AFM in nature. For example, an exchange bias has been reported in PCMO nanoparticles due to the FM and AFM/CO interfacing.[6] Thus, the coexistence of different physical properties due to size reduction is of interest to the general research on complex oxides.

Recently, the research interest on inorganic colloidal nanoparticles has moved to more complex structures, such as anisotropically shaped particles and branched objects.[8] A peculiar reason is that the complex nanostructures further unveils many significant changes over the bulk properties. In the nanocrystalline perovskite manganites, such magnetic properties changes seem to be influenced by the reduced grain size, offering an exciting room when playing with 3-dimensional (3D) nanocomposites. For example, it was shown that vertical interfacing of different materials alters the structural and physical properties in composites. Using this idea, we approached the combination of bottom-up nanomaterial growth and pulsed laser deposition technique to design the composites at lower dimension and also to interface them vertically.[9] Our previous study on nanocrystalline $La_{0.7}Sr_{0.3}MnO_3$ (LSMO) interfacing with MgO nanowire array grown on MgO substrates has indeed shown an enhancement of low field magnetoresistance as well as the importance of growing oxides on unconventional templates.[10] The charge-ordered $Pr_{0.5}Ca_{0.5}MnO_3$ compound is even more interesting since its displays, at low temperature, a huge colossal magnetoresistance as a result of phase separation,[11] opening the route for nano-switching devices and magnetoelectronic applications. For these reasons, we have functionalized ZnO nanowires with an oxide. To achieve this, PCMO were grown in 3D nanocomposite form, through vertically interfacing with ZnO nanowire arrays grown on sapphire substrates and its influence on transport properties are presented in this letter.

ZnO nanowire arrays were grown on *a*-plane oriented sapphire substrates by using a vapor transport method at 900 °C in a tube furnace,[12] and PCMO was deposited using pulsed laser deposition technique. One notable advantage of our synthesis is that the crystallinity of the samples can be well adjusted by using different PLD deposition parameters like temperature, energy, frequency, and gas pressure. In this letter, we focus on two samples with different grain sizes to illustrate the result of this facile synthesis. The samples will be mentioned hereafter as PCMO-NW-LG with smaller grain size and PCMO-NW-HG with larger grain size. In addition, for reference, a PCMO thin film was directly deposited on a flat c-plane oriented sapphire substrate and will be mentioned as "PCMO" hereafter. PCMO was deposited at a substrate temperature of 800 °C with 50 mTorr oxygen pressure for PCMO; 700 °C with 200 mTorr oxygen pressure for PCMO-NW-LG and 720 °C with 300 mTorr oxygen pressure for PCMO-NW-HG. Morphological analyzes were carried out using Scanning

Electron Microscope (SEM) Carl ZEISS SUPRA 55. Magnetization (M) as function of the temperature (T) was obtained from a Magnetic Property Measurement System with a superconducting quantum interface device (MPMS Quantum Design, USA) by applying field (H) parallel to substrate surface. Resistivity (ρ) measurements were done by the standard four-probe technique in a Physical Property Measurement System.

The SEM micrograph shows the ZnO nanowire array grown on *a*-plane sapphire (Fig. 1a) with height and diameter around 1.8 μm and 50 nm respectively. After PCMO deposition, the micrograph shows the growth of PCMO grains (Fig. 1b). In PCMO-NW-LG (Fig. 1b) the diameter of nanowire increases to around 140 nm with PCMO having a grain size around 70 nm whereas in PCMO-NW-HG (Fig. 1c) the PCMO is fully covering the nanowire array with grain size around 100 nm and the nanowire diameter increases to around 1 μm. The reference PCMO (Fig. 1d) shows the grain size around 100 nm. The PCMO films both on flat $Al_2O_3$ substrate and ZnO NR/*a*-$Al_2O_3$ grown along (200) direction (not shown). It is important to mention that the substrate temperature required for growing PCMO on ZnO NR/*a*-$Al_2O_3$ is considerably lower than when depositing directly on $Al_2O_3$ substrate which can be beneficial for devices processing.

M(T) curve PCMO-NW-LG (Fig. 2a) shows a broad peak around 240 K, in agreement with the CO temperature. Furthermore there is a small jump near 110 K indicating the presence FM cluster glass state transition ($T_{CG}$) as evidenced in the PCMO nanoparticles.[6] In literature, the AFM and CO states are reported to be suppressed by reducing the grain size, and can even disappear below 40 nm.[6] Further a reentrant spin glass state has been reported in PCMO with grain size of 300 nm. In addition, the FM cluster glass state transition appears below 150 nm which has been attributed to the reappearance of FM double exchange interaction. Similarly, PCMO-NW-HG sample (Fig. 2b) shows the CO transition at 240 K. The FM cluster glass transition seems however to be vanished probably due to the increased grain size. The derivative curve (see the inset of Fig. 3b) shows a FM cluster glass transition at 110 K for PCMO-NW-LG which has not observed in PCMO-NW-HG, while the CO transition is more pronounced in PCMO-NW-HG than in PCMO-NW-LG, suggesting a more robust CO state with the larger grains size.

The ρ(T) of reference PCMO sample (inset in Fig. 3) does not display a typical CO transition anomaly probably due to strain.[5] In the presence of a 7 T magnetic field, there is a clear change in the resistivity below CO transition, i.e. the resistivity is reduced in presence of magnetic field below CO temperature, a typical behavior of CO compounds. It is important to mention that the reference PCMO sample does not show a strong hysteresis behavior as observed below the CO temperature for PCMO grown on $SrTiO_3$ substrates.[5] Prellier *et al.* reported that substrate-induced strains and thickness influence the stability of CO state.[5] The absence of hysteresis behavior in the as-grown PCMO sample may be due to the low thickness of the film, poor grain connectivity or the large tensile strain due to the $Al_2O_3$ substrate. Fig 4 displays the ρ(T) of PCMO-NW-LG and PCMO-NW-HG. The curves are rather different from that of PCMO film. Both PCMO-NW-LG and PCMO-NW-HG samples show slope change below 150 K and again below 25 K. Below 25 K, the resistance of PCMO-NW-LG sample increases more rapidly than PCMO-NW-HG which can be attributed to the grain scattering and/or the presence of FM cluster glass. Between 25 and 140 K, in presence of magnetic field, PCMO-NW-HG sample shows a small decrease in resistivity which is not observed in PCMO-NW-LG sample. Interestingly, the PCMO-NW-LG and PCMO-NW-HG sample show an increase below 15 K for a high magnetic field, indicating a positive magnetoresistance (MR), which is very different for PCMO itself.

MR(H) curves are shown in Fig. 4. The reference PCMO (inset in Fig. 4a) exhibits negative MR similar to that of bulk and thin films except the absence of hysteresis behavior which was previously discussed and attributed to the robustness of the CO state. Similarly PCMO-NW-LG (Fig. 4a) sample shows negative MR above 50 K. At 25 K, it shows a saturated behavior at high applied magnetic field. Further at 15 K, there is a switch in the MR direction above 8T as the resistance starts increasing above this high magnetic field value. It becomes more evident for the measurement at 10 K where a clear switch from negative to positive MR is observed. At this temperature, resistance starts increasing above 5T. The magnetic field dependent resistivity at different temperatures indicated that, as the temperature decreases, the magnetic field required to switch the MR behavior decreases. The observation of the positive magnetoresistance has already been reported and attributed to the coexistence of ferromagnetic cluster and antiferromagnetic phase, i.e. a phase separation scenario.[13] PCMO-NW-HG sample with larger grain size also shows such intriguing positive magnetoresistance behavior although there is no evidence of FM cluster glass. Nevertheless, the decrease in grain size is known to increase the resistance of the sample whereas our PCMO-NW-LG sample shows lower resistance compare to PCMO-NW-HG in the whole temperature range. It is also important to note that the diameter of nanowire increases to 1 μm for PCMO-NW-HG after PCMO deposition. Hence, we believe that there could be more percent of FM nanocluster phase in PCMO-NW-HG which can explain the positive MR even with low magnetic fields comparing to PCMO-NW-HG. Furthermore PCMO-NW-HG shows the hysteresis loop due to the melting of CO at high magnetic field and the magnetoresistance value increases compare to that of PCMO-NW-LG (Fig. 4b). In addition, the MR curve at 150 K has an intriguing asymmetric nature. Similar asymmetric curve has been observed in Cr-doped PCMO while measuring MR after cooling with magnetic field and it was reported to be due to the coexistence of FM nanoclusters in AFM matrix.[14] Thus, we suggest that a magnetic anisotropic behavior leading to unambiguous asymmetric MR curve occurs in our PCMO grown on ZnO nanowire arrays.

In conclusion, charge-ordered $Pr_{0.5}Ca_{0.5}MnO_3$ was grown on ZnO nanowires. Depending on the grain size of PCMO, the stability of charge-ordering state can be tuned, but more importantly the sign of the magnetoresistance is inverted at low temperatures. We have attributed such features to the coexistence of ferromagnetic clusters on the surface and antiferromagnetic phase in the core of the grains. This work suggests that such a process of growing oxide networks on nanowires can be readily extended to other transition-metal-oxide nanomaterials, and open the doors towards tailoring their functionalities of nanowires, particularly used for magnetic random access nano-devices.

The authors RVKM and WP acknowledge CEFIPRA/IFPACR for their financial support (#3908-1). Work is done in the frame of the Laboratoire Franco-Indien pour la Cooperation Scientifique (LAFICS), and the MERLION program.

**References**

1) W. Prellier, P. Lecoeur, and B. Mercey, J. Phys.: Condens. Matter 13, R915 (2001); A. P. Ramirez, ibid. 9, 8171 (1997).

2) W. Prellier, Amlan Biswas, M. Rajeswari, T. Venkatesan, and R. L. Greene Appl. Phys. Lett. 75, 397 (1999)

3) H. Kuwahara, Y. Tomioka, A. Asamitsu, Y. Moritomo and Y. Tokura, Science 270, 961 (1995)

4) M. Tokunaga, N. Miura, Y. Tomioka and Y. Tokura Phys. Rev. B 57, 5259 (1998)

5) W. Prellier, A. M. Haghiri-Gosnet, B. Mercey, Ph. Lecoeur, M. Hervieu, Ch. Simon, and B. Raveau, Appl. Phys. Lett. 77, 1023 (2000).

6) T. Zhang and M. Dressel, Phys. Rev. B 80, 014435 (2009)

7) A. Sundaresan, R. Bhargavi, N. Rangarajan, U. Siddesh and C. N. R. Rao, Phy. Rev. B 74, 161306R (2006).

8) T. Wu and J. F. Mitchell, Phys. Rev. B 74, 214423 (2006); A. Marcu, T. Yanagida, K. Nagashima, K. Oka, H. Tanaka, and T. Kawai, Appl. Phys. Lett. 92, 173119 (2008).

9) Z. Zhang, Y. H. Sun, Y. G. Zhao, G. P. Li, and T. Wu, Appl. Phys. Lett. 92, 103113 (2008).

10) Z. Zhang, R. Ranjith, B. T. Xie, L. You, L. M. Wong, S. J. Wang, J. L. Wang, W. Prellier, Y. G. Zhao and T. Wu, Appl. Phys. Lett. 96, 222501 (2010).

11) S. De Brion, G. Storgh, G. Chouteau, A. Janossy, W. Prellier ; E. Rauwel Buzin, Eur. Phys. J. B 33 413 (2003).

12) D. L. Guo, X. Huang, G. Z. Xing, Z. Zhang, G. P. Li, M. He, H. Zhang, H. Chen and T. Wu, Phys. Rev. B 83, 045403 (2010)

13) R. Mahendiran, R. Mahesh, R. Gundakaram, A. K. Raychaudhuri and C. N. R. Rao, J. Phys.: Condens. Matter. 8, L455 (1996).

14) R. Mahendiran, A. Maignan, M. Hervieu, C. Martin, and B. Raveau, Appl. Phys. Lett. 77, 1517 (2000).

**Figure Captions**

Figure 1: SEM micrographs of (a) as grown ZnO nanowire arrays on $a$-$Al_2O_3$ (b) PCMO-NW-LG sample (70 nm grain size) (c) PCMO-NW-HG sample (100 nm grain size) and (d) PCMO deposited on flat $Al_2O_3$ substrate.

Figure 2: Temperature dependent magnetic measurement plot of (a) PCMO-NW-LG and (b) PCMO-NW-HG with an applied magnetic field of 1 kOe. Inset figure shows the derivative curve of temperature dependent magnetic moment.

Figure 3: Temperature dependent resistivity curve of PCMO-NW-LG and PCMO-NW-HG with and without magnetic field. Inset shows the measurement on PCMO

Figure 4: Magnetic field dependent resistivity curves of (a) PCMO-NW-LG and (b) PCMO-NW-HG at different temperatures. Inset shows the measurement on PCMO

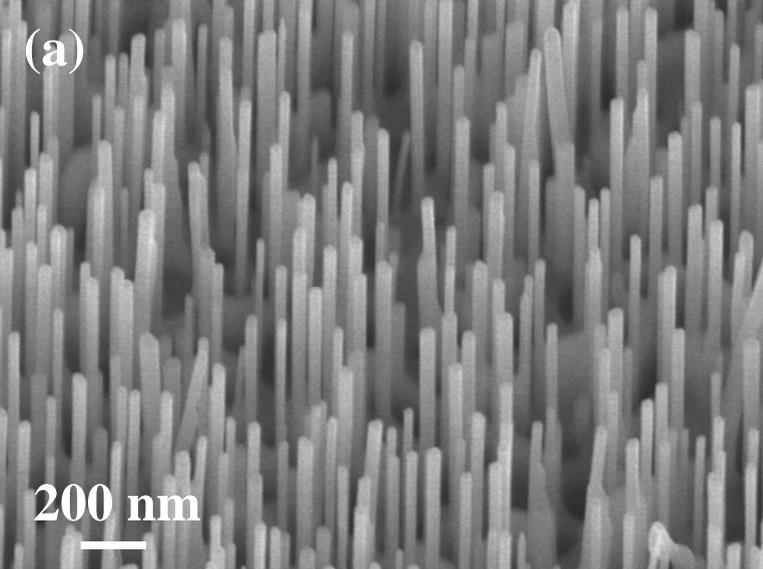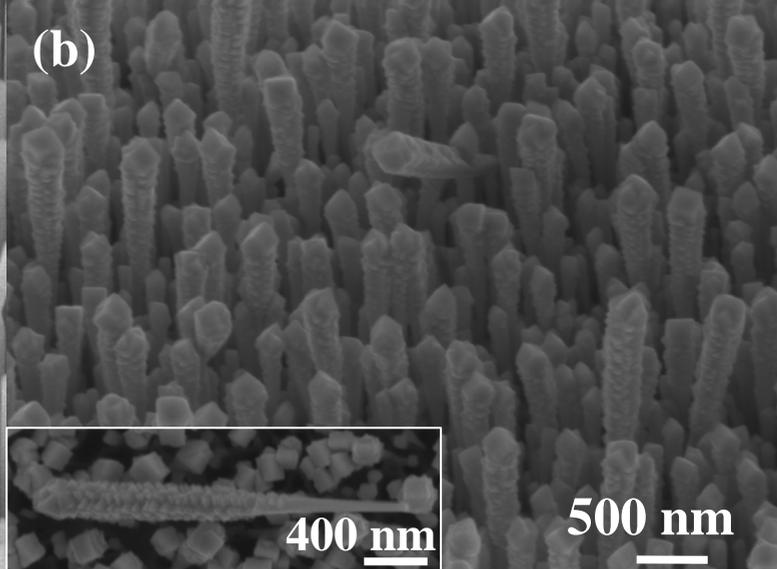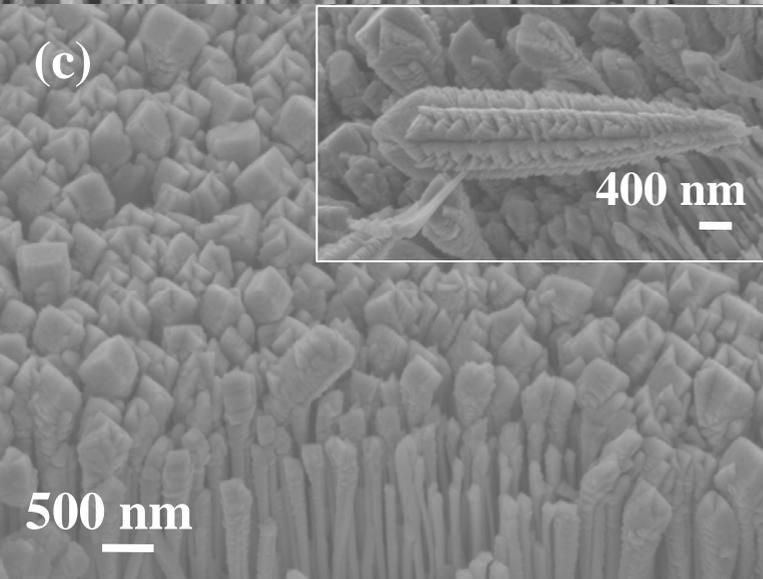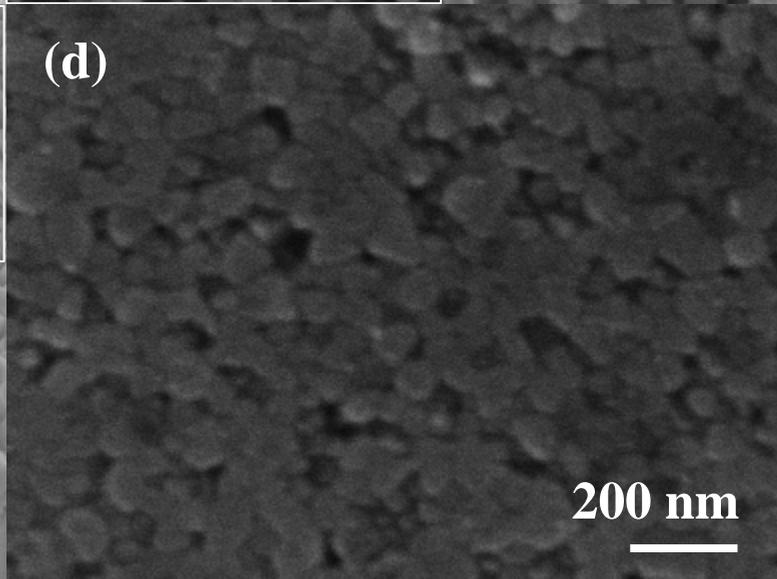

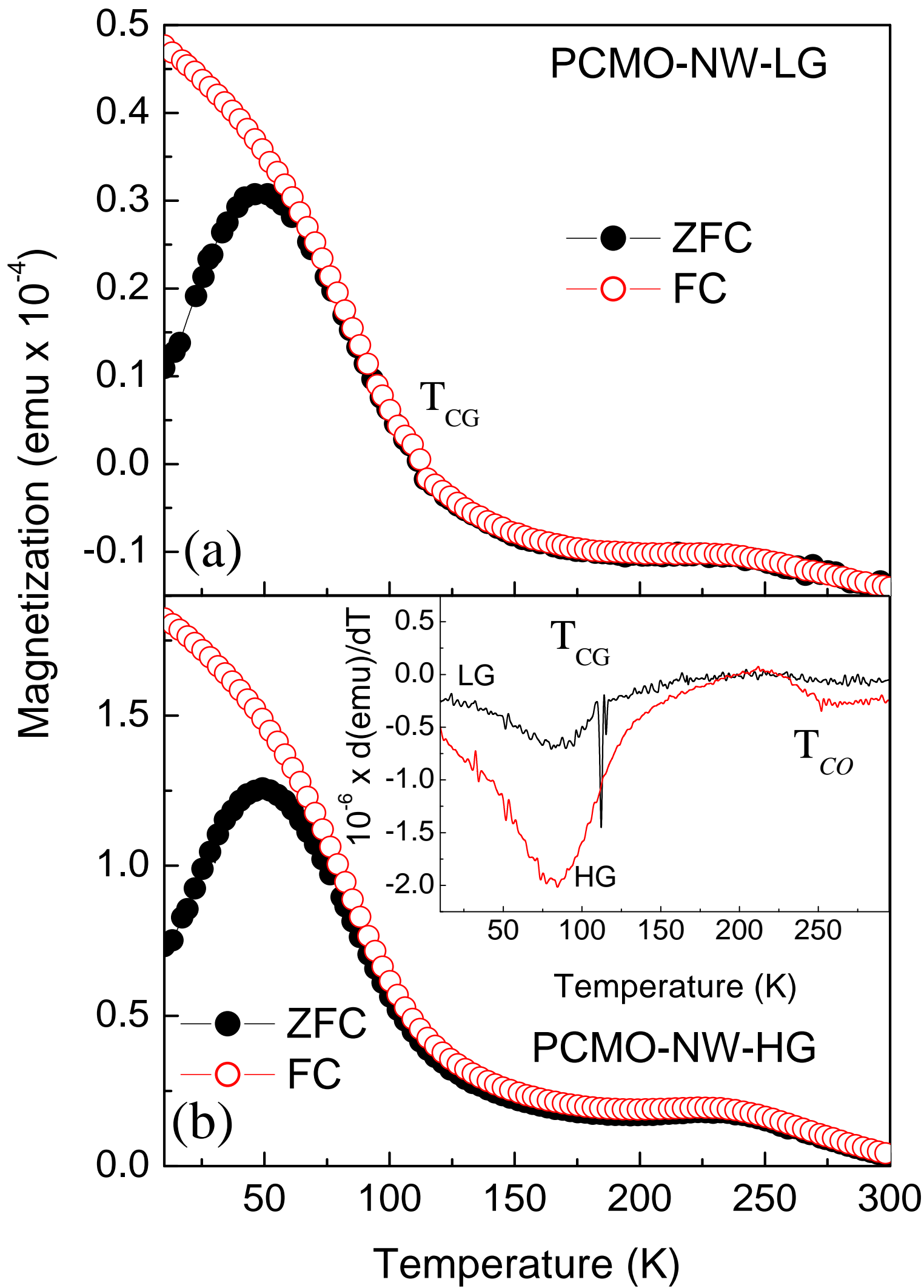

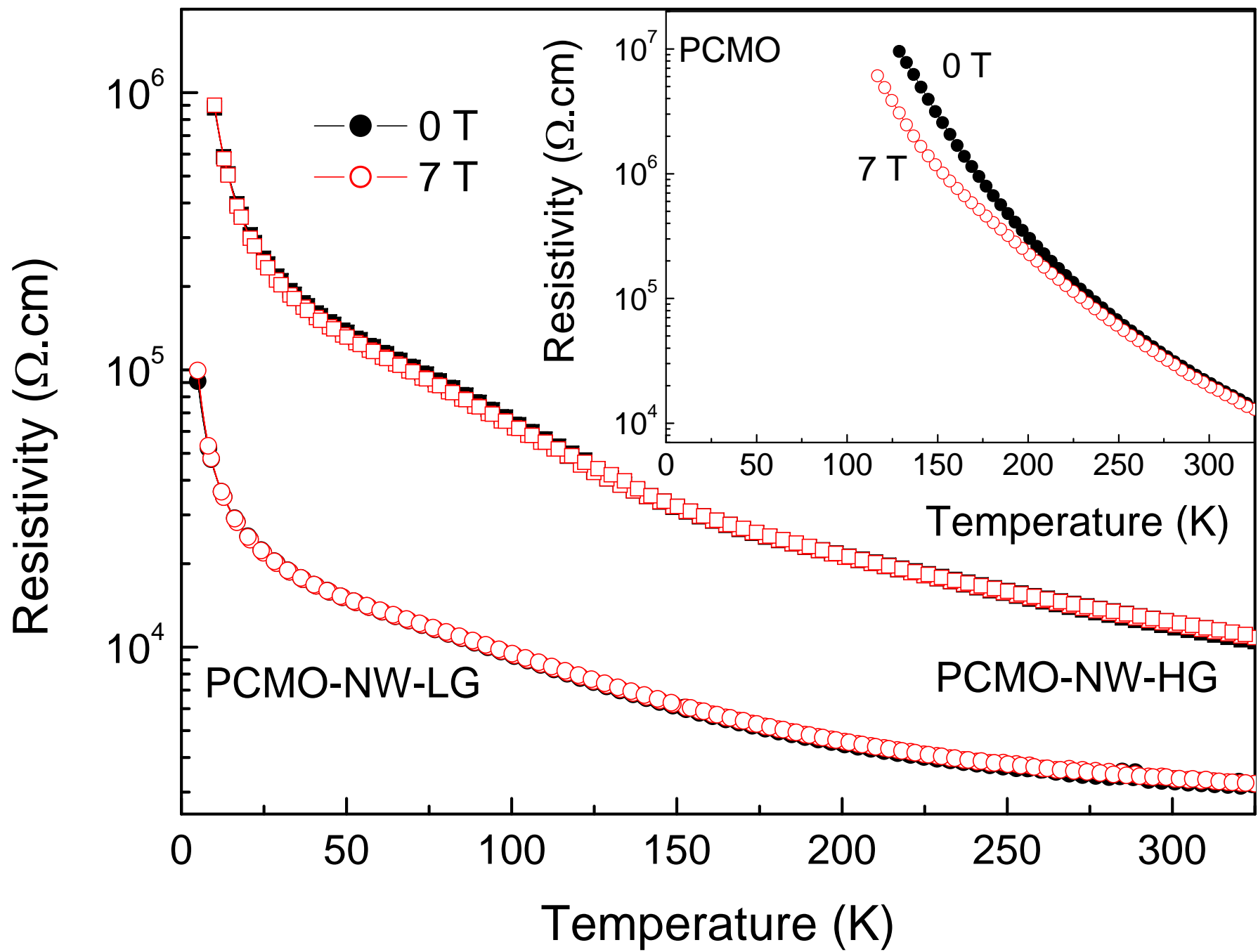

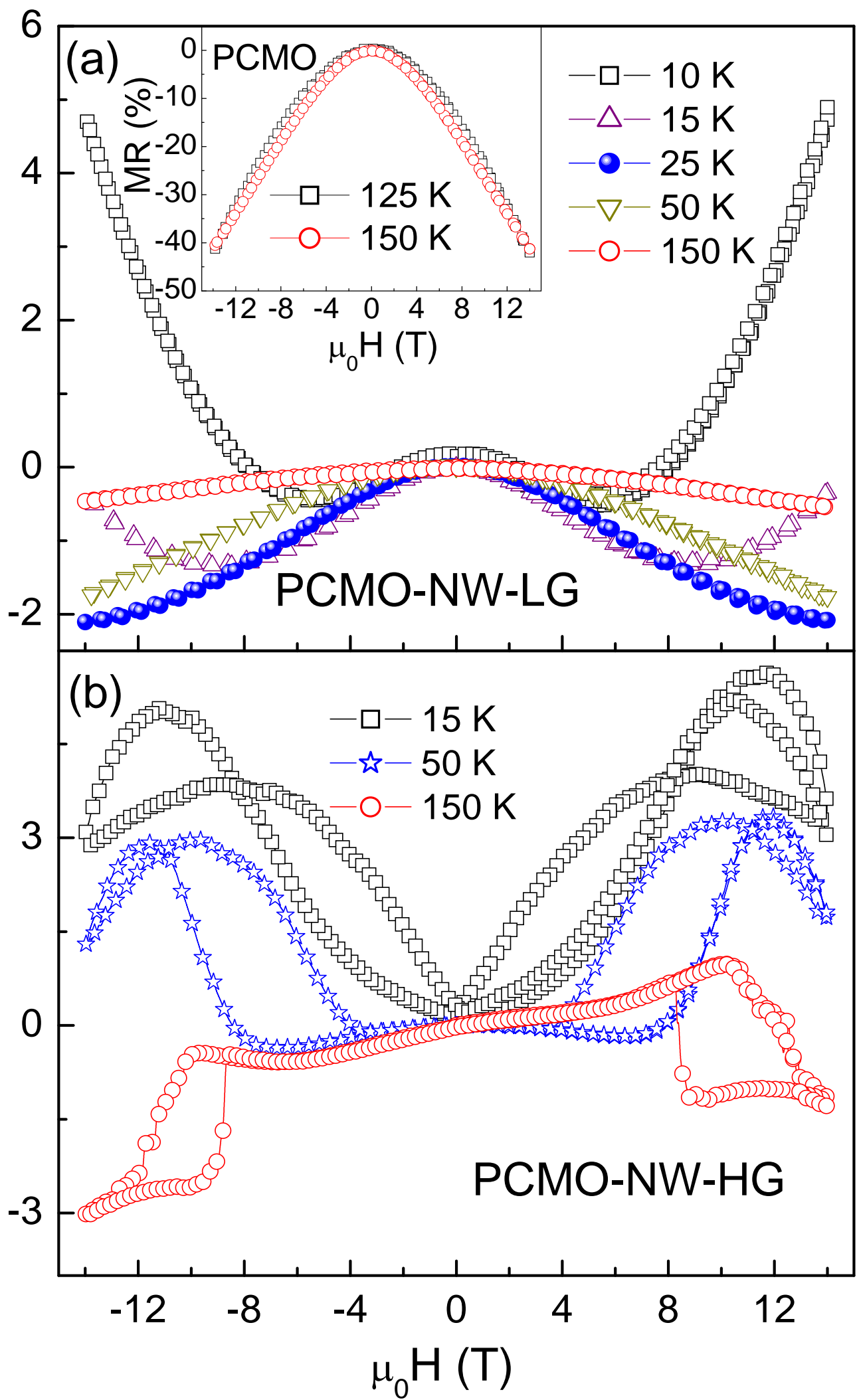